# Physiological Sex-Specific Haematocrit Has Minimal Effect on Coronary Computational Haemodynamics: Modelling Implications for Blood Rheology

**C. Shen[1], M. Zhang[1,2], T. Shalaby[1], C. S. McLachlan[1,2], S. Beier[1]**

[1]*School of Mechanical and Manufacturing Engineering, University of New South Wales, Sydney NSW 2052, Australia*

[2]*Health Hub, Cardiovascular translational group, Torrens University Australia, Surry Hills, Sydney, Australia*

**Abstract**

**Background:**

Haematocrit influences blood viscosity and may affect coronary computational fluid dynamics (CFD). However, previous studies examined broad or pathological haematocrit ranges, and it remains unclear whether female-specific haematocrit variations within the physiological range produce meaningful changes in coronary haemodynamics.

**Methods:**

Fifteen female coronaries were analysed, including healthy arteries and corresponding diseased models with mild, moderate and severe stenosis. A haematocrit-dependent Carreau–Yasuda model was developed using shape-preserving piecewise cubic Hermite interpolation of haematocrit-specific viscosity parameters. Coronary CFD simulations were performed using the standard rheology model and a female-specific haematocrit-based model (40%). Time-averaged endothelial shear stress (TAESS), endothelial shear stress gradient (ESSG), temporal shear variation index (TSVI), helicity intensity, and low/high TAESS exposure were compared across coronary trees, arterial segments, bifurcations, stenosed vessels and corresponding narrowed regions.

**Results:**

The female-specific model produced statistically significant differences from the standard model across all haemodynamic metrics and coronary regions ($p<0.05$). However, the absolute differences were small, indicating a limited haemodynamic impact. Bland–Altman analysis showed narrow biases and limits of agreement. Linear regression demonstrated significant associations between inter-model differences and haemodynamic magnitude for TAESS, ESSG, helicity intensity, and adverse TAESS exposure, but the slopes were consistently small. Similar findings were observed in stenosed arteries, where both models captured comparable flow disturbances across stenosis severities.

**Conclusions:**

Female-specific haematocrit variation within the physiological range is computationally detectable but haemodynamically negligible in coronary CFD. A standard rheology model is therefore likely sufficient for most coronary CFD studies, while personalised haematocrit modelling is more relevant for patients with abnormal haematocrit or rheology-focused studies.

## 1. Introduction

Blood flow exhibits shear-thinning, non-Newtonian behaviour, whereby its viscosity decreases as shear rate increases. In the coronary arteries, this shear-dependent rheology acts together with complex anatomical features, such as bifurcations, curved segments, and stenoses, to shape disturbed flow patterns and near-wall haemodynamic conditions [1, 2]. These local haemodynamic environments have been associated with the development of atherosclerotic plaques through their effects on endothelial functions [3].

Among the determinants of blood viscosity, haematocrit (Hct) characterises the volume fraction of red blood cells and influences both the magnitude and shear-rate dependence of blood viscosity [4, 5]. Haematocrit is clinically relevant, with both low and high levels associated with critically increased mortality [6, 7]. Low haematocrit is commonly associated with anaemia [17], whereas high haematocrit can occur in conditions such as polycythaemia vera, for which diagnostic thresholds include haematocrit values above 49% in males and 48% in females [18], with treatment recommendations aiming to maintain haematocrit below 45% [19]. Haematocrit values outside the normal range are clinically associated with vascular dysfunction in males [16]. As a result, it remains unclear whether haematocrit-related differences observed across broad or pathological ranges translate into meaningful changes in coronary haemodynamics within normal physiological ranges.

Computational fluid dynamics (CFD) remains a key method for investigating coronary haemodynamics, and naturally lends itself to haematocrit studies. Prior computational works tested ranges of 10-65% [8-10], linking higher haematocrit to increased TAESS magnitude [11], location and exposure to adverse TAESS [8]. When Hct increased, a reduced low TAESS exposure was observed in bifurcation regions [9], whilst in stenosed regions, a rapid increase was noted at the minimal lumen area [10]. It is important to note, however, that these represent abnormal haematocrit ranges rather than the normal physiological ranges typically observed in females (~36-47%) and males (~40-50%) [7, 12-14]. In fact, modest haematocrit variations (less than approximately 10%, i.e. ~31-52% in females and ~35-55% in males) may be partly mitigated by compensatory regulatory mechanisms in the human circulation [15]. More importantly, although haematocrit differs physiologically between females and males [7, 12], this sex-related variation has not been considered in previous coronary CFD studies, despite the established influence of sex on coronary anatomy and haemodynamics [20, 21].

Accordingly, the present study aimed to determine whether incorporating sex-specific haematocrit values within the normal physiological range leads to meaningful changes in coronary haemodynamic metrics, and whether these effects persist in both healthy and narrowed coronary arteries. To address this, we developed a haematocrit-dependent blood rheology framework based on a modified Carreau–Yasuda model and compared standard and sex-specific simulations across multiple coronary regions and haemodynamic factors. This work aims not only to clarify the haemodynamic relevance of sex-specific haematocrit variations, but also to provide practical modelling guidance on when a standard rheology model is sufficient and when patient-specific rheological personalisation may be warranted.

## 2. Methods

### 2.1. Patient Geometries

Fifteen healthy female coronary artery models from the ASOCA dataset and fifteen identical diseased models, with synthetically generated stenotic narrowing, were used in this study. Only female cases were included because the standard Carreau–Yasuda (CY) blood rheology model is based on a representative Hct value of 45% [22], which is consistent with the average physiological range observed in males [12]. Therefore, this study focused on the Hct effects on haemodynamics in a female-specific manner.

The diseased models were generated from the corresponding healthy coronary geometries using the MorphMan package [23], allowing controlled stenosis morphology and more consistent haemodynamic comparison. Stenoses were introduced in the proximal left anterior descending artery (LAD), located two local vessel diameters distal to the left main bifurcation, with a lesion length of two local vessel diameters. Three stenosis severities were investigated: mild (35%), moderate (55%), and severe (75%), consistent with CAD-RADS classification categories [24] to further evaluate the influence of Hct across different disease severities.

### 2.2. Haematocrit-dependent Rheology Model

The Carreau–Yasuda (CY) non-Newtonian rheology model was used to describe blood viscosity [25]:

$$\mu(\dot{\gamma}) = \mu_\infty + (\mu_0 - \mu_\infty)[1 + (\lambda\dot{\gamma})^a]^{\frac{n-1}{a}}$$

$\mu(\dot{\gamma})$ is the apparent viscosity ( $\mathrm{Pa \cdot s}$), $\dot{\gamma}$ is the shear rate ($s^{-1}$), $\mu_0$ represents the zero-shear viscosity, and $\mu_\infty$ represents the infinite-shear viscosity. The parameters $\lambda, a, n$ represent the relaxation time constant, Yasuda exponent, and power-law index, respectively.

Although the CY model captures the shear-thinning behaviour of blood, it does not explicitly account for viscosity changes caused by haematocrit variations. In this work, a modified haematocrit-dependent CY model was developed to account for the effect of haematocrit on blood viscosity, following and modifying the methods in previous literature [26]. Each Carreau–Yasuda parameter was treated as a haematocrit-dependent quantity. The parameter set was therefore written as

$$\theta(\phi) = \big(\mu_0(\phi), \mu_\infty(\phi), \lambda(\phi), a(\phi), n(\phi)\big)$$

where the continuous relationship between each parameter and haematocrit $\phi$ was constructed using shape-preserving piecewise cubic Hermite interpolation (PCHIP) [27]. This allowed the model to estimate a patient- or case-specific viscosity curve for a given haematocrit value while preserving the fitted trends between the available experimental haematocrit levels.

Experimental whole-blood viscosity datasets at different Hct levels were adopted from previous literature [4]. The 25%, 65%, and 95% haematocrit datasets were used for model fitting, while the 45% dataset was withheld for validation. For each training haematocrit level, five CY parameters $(\mu_0, \mu_\infty, \lambda, a, n)$ were estimated using nonlinear least-squares optimisation in logarithmic viscosity space to balance errors across the low- and high-shear-rate regions.

After the individual haematocrit-specific fits were obtained, the fitted parameter values were used as anchor points for the PCHIP interpolation. Parameters $(\mu_0, \mu_\infty, \lambda, a)$ were interpolated after log transformation to ensure that the interpolated values remained positive, while the power-law index $n$ was interpolated using a logit transformation to maintain $0 < n < 1$, consistent with the shear-thinning behaviour of blood. The resulting interpolated parameter set $\theta(\phi)$ was then substituted into the Carreau–Yasuda equation to calculate blood viscosity for the target haematocrit values used in the simulations. The fitting and validation results are shown in Figure 1.

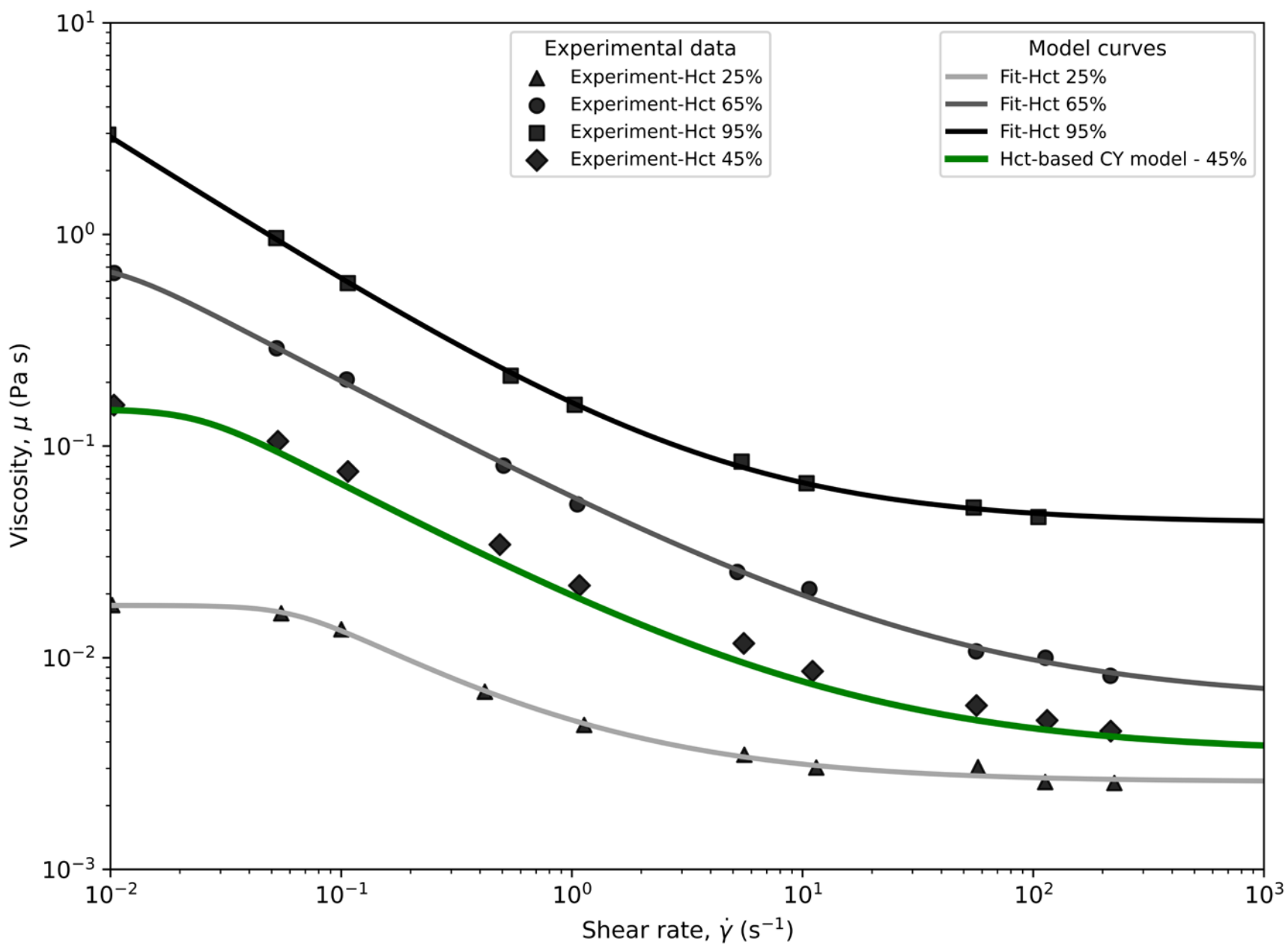

*Figure 1. Fitting and validation of the haematocrit (Hct)-dependent Carreau–Yasuda (CY) model. Experimental viscosity–shear rate data from Chien et al. are shown as black markers. The 25%, 65%, and 95% haematocrit datasets were used for model fitting, and their fitted CY curves are shown in grey, dark grey, and black, respectively. The 45% haematocrit dataset was withheld from fitting and used for validation. The green curve shows the haematocrit-dependent CY model prediction at 45% haematocrit obtained from PCHIP-interpolated parameters.*

### 2.3. Computational Fluid Dynamics

All coronary artery simulations were performed using ANSYS CFX (ANSYS Inc., Canonsburg, PA, USA), while mesh generation was conducted using ICEM-CFD within the ANSYS package (version 2023R1). A mesh sensitivity analysis was performed prior to the simulations to ensure numerical accuracy and solution independence.

Blood flow was assumed to be incompressible and laminar. For the female-specific Carreau–Yasuda model, haematocrit was set to 40%, representing a population-level female value reported in the literature. Specifically, this value was calculated as a weighted mean across reported female haematocrit data to provide a representative estimate for the female population. As patient-specific flow measurements were unavailable, inlet boundary conditions were prescribed using an allometric scaling relationship [28, 29]. The cycle-averaged inlet flow rate was calculated as:

$$Q = 1.43d^{2.55}$$

where $Q$ is the cycle-averaged flow rate, and $d$ is the mean diameter of the left main coronary artery. Distal outlet flow distribution was determined using a flow-splitting relationship based on vessel diameter [28]:

$$\frac{Q_{sb}}{Q_{mb}} = (\frac{d_{sb}}{d_{mb}})^{2.27}$$

where $Q_{sb}$ and $Q_{mb}$ represent the flow rates in the side branch and main branch, respectively, while $d_{sb}$ and $d_{mb}$ are their corresponding mean diameters. Vessel walls were assumed rigid with a no-slip boundary condition applied at the lumen surface.

A steady-state simulation was first performed for each model to initialise the transient simulations. Transient simulations covering four cardiac cycles were conducted, and haemodynamic results from the final cycle were extracted for analysis to minimise start-up transient effects.

ESS was quantified using its time-averaged value, TAESS and the time-averaged ESS gradient (ESSG) on the luminal surface. Temporal shear variation index (TSVI) was evaluated to characterise cyclic endothelial contraction and expansion behaviour [30, 31]. Time-averaged absolute helicity intensity $h_2$ was measured regarding the rotational movement of the blood flow [32, 33]. Adverse haemodynamic exposure was evaluated using established TAESS thresholds of 0.5 Pa for low TAESS, and 4.71 Pa for high TAESS [1]. The proportion of lumen surface area exposed to these adverse haemodynamic conditions was calculated and reported as lowTAESS and highTAESS, respectively.

### 2.4. Statistical Analysis

Statistical analyses were performed using the Python Statsmodels package [34]. The influence of Hct on coronary haemodynamics was evaluated across multiple anatomical regions, including the entire left and right coronary artery trees, individual vessels (left main coronary artery [LM], LAD, and left circumflex artery [LCx]), and bifurcation regions. For stenosed geometries, additional analyses were performed in the diseased vessel (LAD in this work) and within the stenosed region, defined as 1.5 mm proximal and distal to the lesion.

Haemodynamic metrics obtained with the female-specific (FS) and standard (SD) CY models were compared using paired statistical analysis. Data normality was assessed using the Shapiro–Wilk test. For normally distributed data, paired t-tests were applied, whereas non-normally distributed data were analysed using the Wilcoxon signed-rank test. Statistical significance was defined at $p<0.05$, with false discovery rate (FDR) correction performed using the Benjamini–Hochberg method to account for multiple comparisons.

For normally distributed variables, results were reported as mean ± standard deviation, while non-normally distributed variables were reported as median and interquartile range. Agreement between SS and SD models was further evaluated using Bland–Altman analysis, including calculation of bias and 95% limits of agreement. Linear regression analyses were additionally performed to assess the relationship between inter-model differences and mean haemodynamic values.

## 3. Results

### 3.1. Effect of Haematocrit in Healthy Coronary Arteries

Figure 1 compares haemodynamic metrics obtained using the standard and female-specific Carreau–Yasuda viscosity models across bifurcations, arterial segments, and entire coronary artery trees. Statistically significant differences were observed between the two models for all haemodynamic metrics ($p < 0.05$). However, the overall magnitude of differences remained small across all analysed coronary regions.

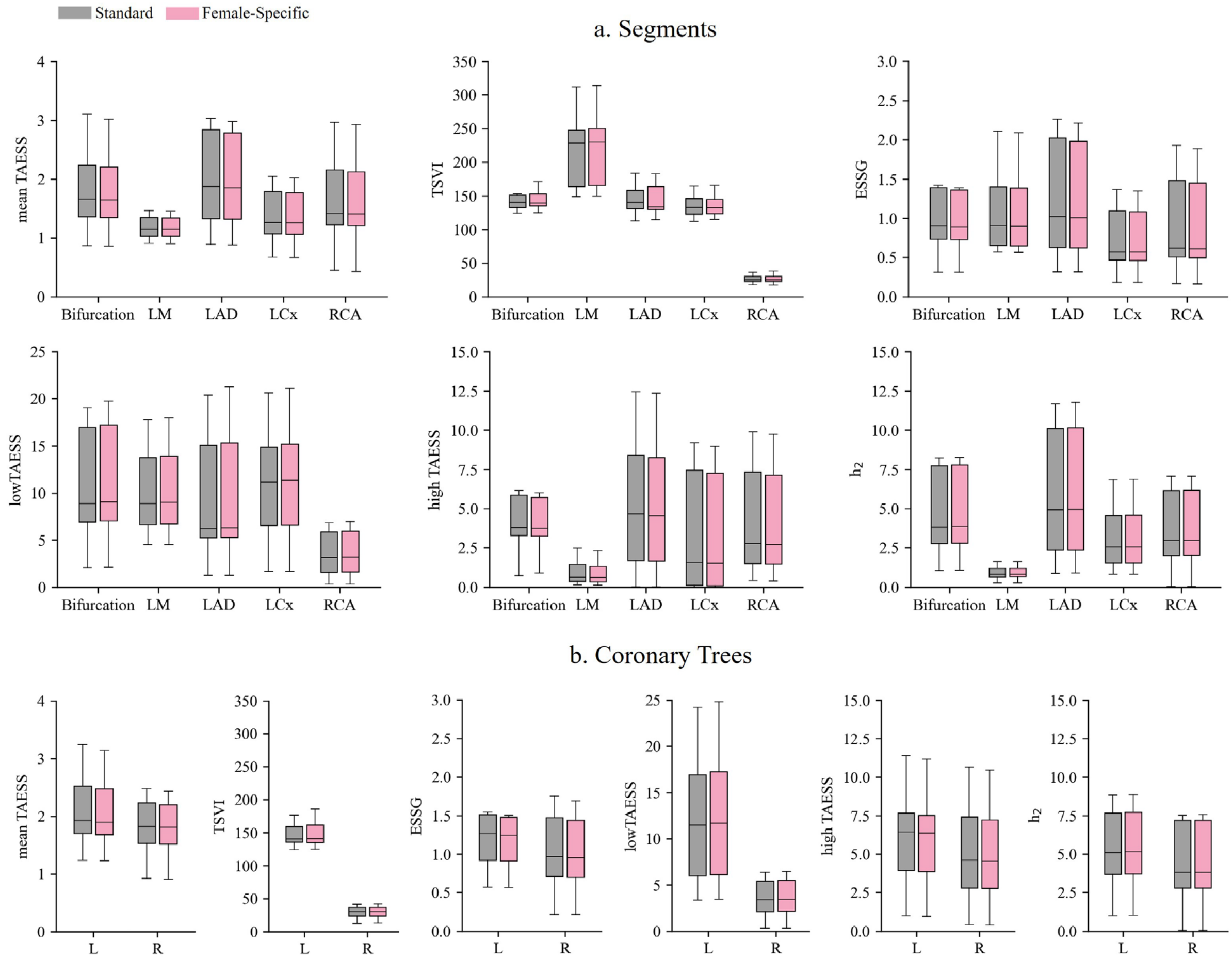

*Figure 2. Comparison of haemodynamic factors simulated using standard and female-specific Carreau–Yasuda blood viscosity models across coronary regions. a. Segments include bifurcation, left main coronary artery, LM, left anterior descending artery, LAD, left circumflex artery, LCx, and right coronary artery main branch, RCA. b. Coronary artery trees include the left and right coronary artery trees (L and R, respectively).*

Bland–Altman analysis demonstrated consistent agreement between the two viscosity models across bifurcations, arterial segments, and entire coronary trees (Figure 2). In female-specific simulations, mean TAESS and ESSG demonstrated a decreasing trend with increasing haemodynamic magnitude, whereas lowTAESS and $h_2$ showed increasing differences at larger values. Despite these relationships, the overall biases and limits of agreement remained narrow for most haemodynamic metrics. No trend was observed in TSVI.

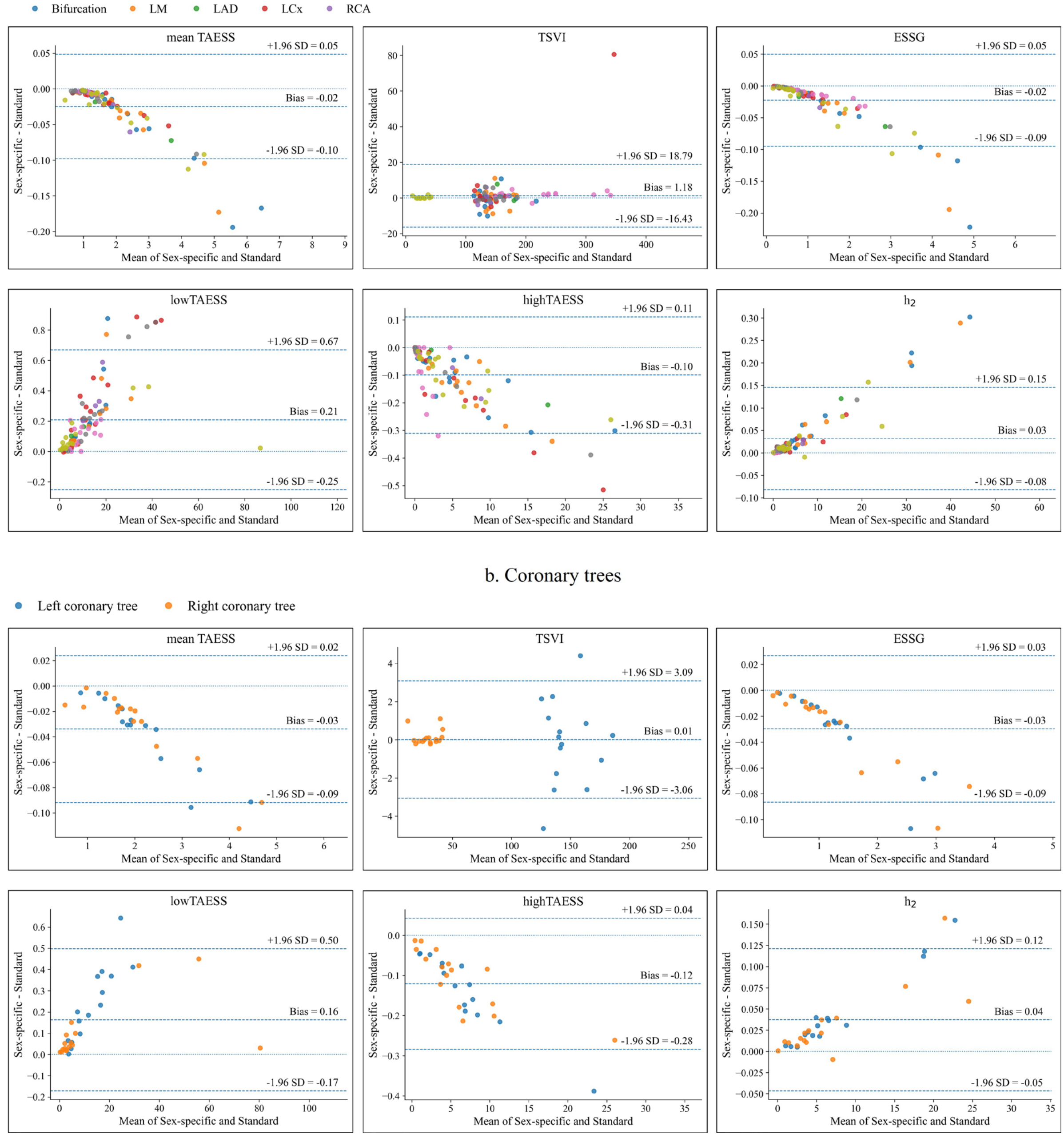

*Figure 3. Bland–Altman plots comparing haemodynamic metrics simulated using the generic and female-specific Carreau–Yasuda viscosity models across arterial segments. The x-axis represents the mean of the two models, and the y-axis shows their difference (female-specific-standard). Dashed lines indicate the mean bias, and the limits of agreement (±1.96 SD).*

Linear regression analysis of inter-model haemodynamic differences is summarised in Table 1. Differences in mean TAESS, ESSG, low/highTAESS, and $h_2$ between two CY models showed strong correlations with haemodynamic magnitude across all regions. Although these haemodynamic metrics demonstrated strong statistical correlations, the regression slopes were consistently small, indicating that the magnitude of model-induced haemodynamic differences remained limited even at higher haemodynamic values. Mean TSVI showed no correlation across all anatomical regions.

*Table 1. Linear regression analysis of the absolute differences between haemodynamics factors simulated using the standard and female-specific Carreau–Yasuda viscosity models across bifurcations, arterial segments, and entire coronary trees.*

| Region | Metric | Slope | $R^2$ | p-value |
|---|---|---|---|---|
| **Bifurcation** | Mean TAESS | 0.03 | 0.899 | <0.001 |
| | Mean TSVI | 0.00 | 0.000 | 0.990 |
| | ESSG | 0.03 | 0.883 | <0.001 |
| | Low TAESS | 0.02 | 0.830 | <0.001 |
| | High TAESS | 0.02 | 0.871 | <0.001 |
| | $h_2$ | 0.01 | 0.975 | <0.001 |
| **Arterial segments** | Mean TAESS | 0.03 | 0.903 | <0.001 |
| | Mean TSVI | 0.05 | 0.130 | 0.004 |
| | ESSG | 0.03 | 0.831 | <0.001 |
| | Low TAESS | 0.01 | 0.242 | <0.001 |
| | High TAESS | 0.01 | 0.590 | <0.001 |
| | $h_2$ | 0.01 | 0.906 | <0.001 |
| **Entire Trees** | Mean TAESS | 0.03 | 0.864 | <0.001 |
| | Mean TSVI | 0.00 | 0.004 | 0.742 |
| | ESSG | 0.03 | 0.841 | <0.001 |
| | Low TAESS | 0.01 | 0.217 | 0.008 |
| | High TAESS | 0.01 | 0.738 | <0.001 |
| | $h_2$ | 0.01 | 0.784 | <0.001 |

### 3.2. Effect of Haematocrit in Stenosed Coronary Arteries

Figure 3 compares haemodynamic metrics obtained using the standard and female-specific Carreau–Yasuda viscosity models within the stenosed vessel and lesion region. The difference was statistically significant for all haemodynamic metrics between the two models (all $p < 0.05$), however, the magnitude of these differences remained small across both analysed regions.

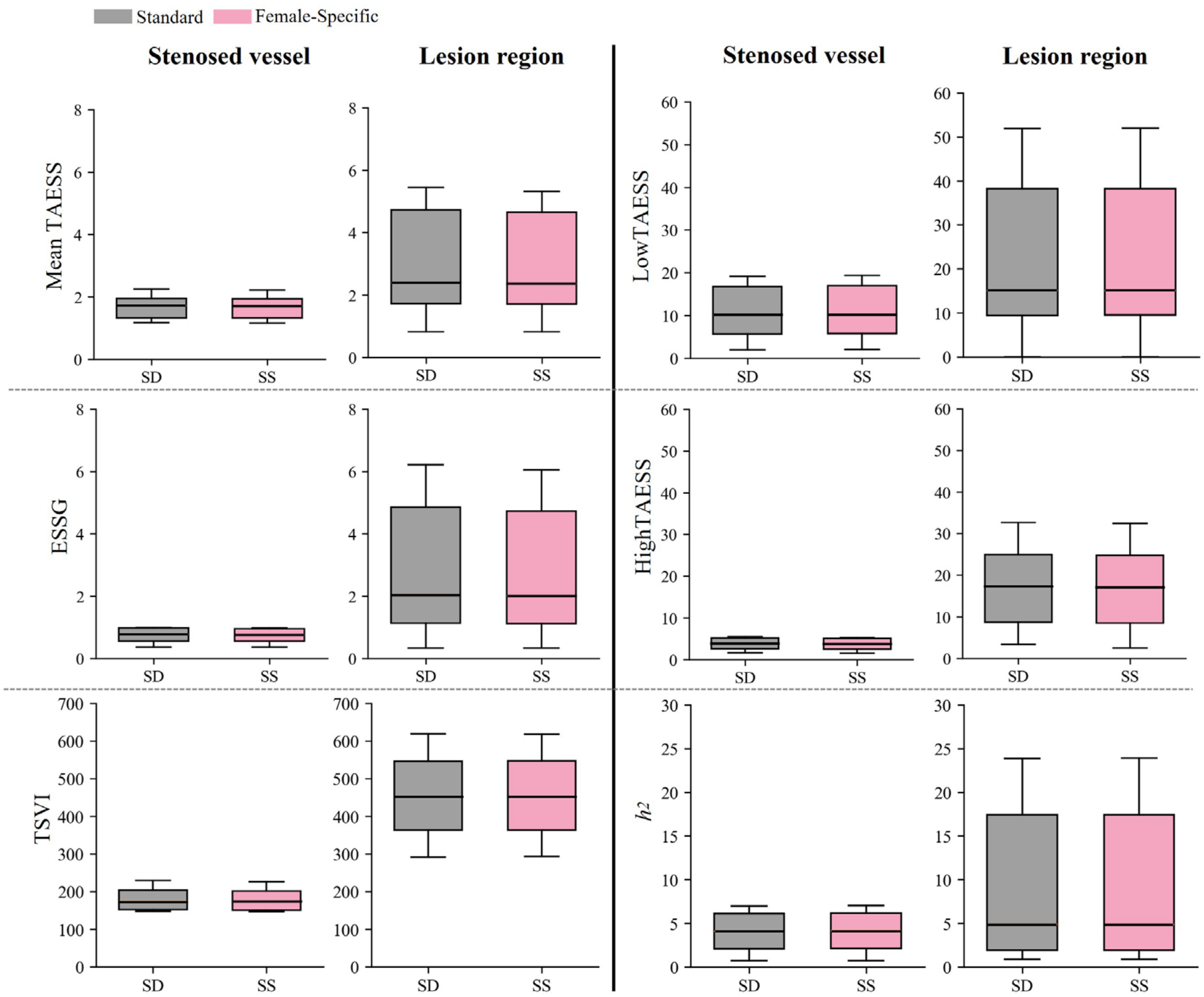

*Figure 4. Comparison of haemodynamic factors simulated using standard and female-specific Carreau–Yasuda blood viscosity models in stenosed vessel and narrowed regions.*

Linear regression analysis showed statistically significant relationships between inter-model haemodynamic differences and haemodynamic magnitude for mean TAESS, ESSG, and $h_2$ in both the stenosed vessel and the lesion region. LowTAESS and highTAESS demonstrated moderate and weak correlations within the stenosed vessel, respectively. Again, the slope was minimal. The TSVI difference between CY models showed no correlation with its magnitude in either region.

*Table 2. Linear regression analysis of the absolute differences between haemodynamics factors simulated using the standard and female-specific Carreau–Yasuda viscosity models in stenosed vessels and narrowed regions.*

| Region | Metric | Slope | $R^2$ | p-value |
|---|---|---|---|---|
| **Stenosed vessel** | Mean TAESS | -0.02 | 0.969 | <0.001 |
| | Mean TSVI | 0.01 | 0.009 | 0.737 |
| | ESSG | -0.02 | 0.993 | <0.001 |
| | Low TAESS | 0.02 | 0.775 | <0.001 |
| | High TAESS | -0.01 | 0.370 | 0.016 |
| | $h_2$ | 0.01 | 0.931 | <0.001 |
| **Lesion region** | Mean TAESS | -0.03 | 0.991 | <0.001 |
| | Mean TSVI | 0.01 | 0.052 | 0.412 |

|  | ESSG | -0.04 | 0.992 | <0.001 |
|---|---|---|---|---|
|  | Low TAESS | 0 | 0.041 | 0.470 |
|  | High TAESS | -0.01 | 0.107 | 0.233 |
|  | $h_2$ | 0 | 0.667 | <0.001 |

Representative haemodynamic patterns across mild, moderate, and severe stenosis severities are shown in Figure 4. As stenosis severity increased, recirculation regions developed distal to the lesion, with low TAESS distribution appearing before and after the stenosed regions. However, these disturbed flow structures were consistently captured by both viscosity models. Velocity streamlines, TAESS contour maps, and adverse TAESS exposure contour maps were identical between the standard and female-specific CY models across all stenosis severities.

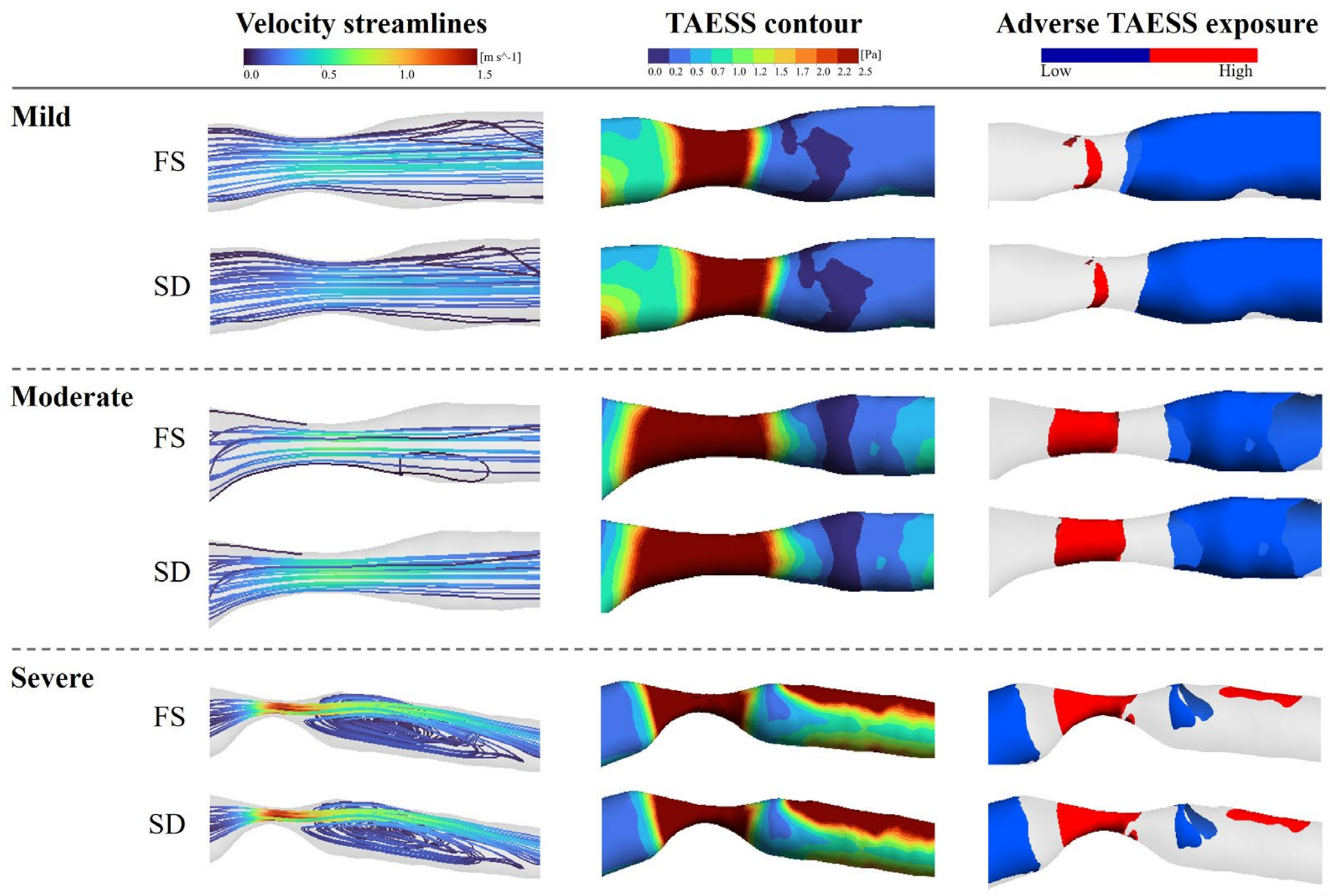

*Figure 5. Representative coronary artery haemodynamic patterns obtained using the standard (SD) and female-specific (FS) Carreau–Yasuda models in arteries with mild, moderate, and severe stenosis. Velocity streamlines, TAESS contours, and adverse TAESS exposure distributions are shown for each case severity. Adverse TAESS exposure indicated the region exposed to low endothelial shear stress (TAESS) <0.5 Pa (blue), and high TAESS > 4.71 Pa (red). Across the three stenosis severities, all the haemodynamic patterns showed identical distributions.*

## 4. Discussion

The main finding of this study is that physiological female-specific differences in haematocrit produced statistically detectable but haemodynamically negligible changes in both healthy and stenosed coronary arteries. Across arterial segments, bifurcations, and stenosed regions, the standard and female-specific rheology models yielded comparable haemodynamics in healthy coronaries and captured flow patterns in diseased arteries with mild, moderate, and severe stenosis. These results indicate that normal physiological sex-specific variation in haematocrit can be detected computationally, but is unlikely to change the haemodynamic interpretation of most coronary blood flow simulations.

Previous studies have suggested that haematocrit can have a substantial impact on coronary haemodynamics [8-10]. In contrast, the present results suggest that this effect becomes limited when haematocrit is restricted to normal physiological, sex-specific values. This is likely due to the much broader haematocrit ranges examined in earlier work (~10-65%), which far exceed the normal ranges for females and males (~36-50%). Another explanation is that, in coronary arteries, local anatomical features play a more dominant role than rheological variation. In particular, curvature, bifurcations and stenosis strongly shape the flow patterns. This is further supported by the lack of TSVI sensitivity to haematocrit variation in this work, as TSVI reflects the endothelium shear stress contraction/expansion variability, and is therefore more strongly governed by local geometry and flow patterns [30, 31]. In severe stenosed cases, both standard and female-specific models captured the typical stenosis-induced jet flows, with elevated TAESS within the narrowed region and low TAESS distributions upstream and downstream of the stenosis. The near-identical haemodynamic patterns produced by the two models suggest that these disturbed flow features were primarily driven by local geometry and lesion severity, rather than by modest haematocrit variations within the physiological range tested here.

Nevertheless, these findings should not be interpreted as evidence that haematocrit is haemodynamically unimportant in all settings. Large haematocrit variations can shift both the overall viscosity level of blood and its oxygen transport capacity. In this work, although the Bland–Altman and linear regression analyses showed only small inter-model differences within the normal female physiological range, they also indicated a consistent magnitude-related trend that aligns with previous studies [8-10]: higher Hct levels tended to increase TAESS magnitude and adverse high-TAESS exposure. This agrees with the broader literature suggesting that abnormally high haematocrit, such as in polycythaemia vera, may increase blood viscosity, coronary vascular resistance and thrombotic risk, which may become particularly relevant in stenosed regions where excessive shear exposure is already present [18, 19]. By contrast, the mechanism associated with low haematocrit may not be mainly driven by induced haemodynamic variations. Rather, in conditions such as anaemia, reduced haemoglobin lowers the oxygen-carrying capacity of blood, limiting myocardial oxygen supply, while compensatory increases in heart rate and stroke volume may increase myocardial oxygen demand [17]. This oxygen supply–demand imbalance may help explain why low haematocrit is associated with adverse outcomes, especially in patients with coronary stenosis or acute coronary syndromes. Therefore, the present findings do not contradict previous studies, instead, they suggest that haematocrit-related effects are minimal within the normal physiological range but may become clinically important when haematocrit falls outside this range through different low- and high-haematocrit mechanisms.

These findings have direct methodological implications for coronary CFD workflows. For most healthy and stenosed coronary simulations with haematocrit values within the physiological range, a standard rheology model may be sufficient. This can simplify model setup, reduce the need for routine haematocrit personalisation, and support more consistent large-scale or population-based simulations. However, personalised haematocrit modelling should still be considered when haematocrit is abnormal or when the clinical question specifically concerns altered blood rheology. Therefore, the practical message is not that haematocrit can always be ignored, but that its inclusion should depend on the physiological or pathological context of the study.

Several limitations should be acknowledged. First, the sample size and set of coronary geometries may limit generalisability across coronary anatomy and disease phenotypes in broader populations. Second, clinically measured patient-specific haematocrit values were not available, so the female-specific rheology conditions were based on physiological reference values rather than paired subject-specific data. Third, the simulations did not incorporate wall motion or vascular compliance, which may affect local haemodynamic patterns. In addition, while this study focused on coronary arteries and physiologically plausible sex-specific haematocrit differences, the conclusions should not be extrapolated to broader pathological haematocrit ranges without further testing. Future work should therefore evaluate patient-specific haematocrit measurements, extend the analysis to pathological states, and investigate whether similar conclusions hold under more complex modelling assumptions. Overall, the present findings support the conclusion that physiological sex-specific haematocrit differences are computationally detectable but haemodynamically negligible in coronary CFD, and that personalised rheology is most likely to add value in abnormal rather than routine physiological settings.

## 4. Conclusion

This study investigated the effect of a female-specific haematocrit-based blood rheology model in 15 female coronary cases, including healthy arteries and corresponding diseased models with mild, moderate and severe stenosis. Haematocrit variation within the normal physiological range produced statistically detectable but haemodynamically negligible changes in coronary haemodynamics. These findings suggest that a standard rheology model is likely to be sufficient and may provide a practical balance between modelling simplicity and haemodynamic accuracy. Personalised haematocrit modelling should therefore be considered for patients with abnormal haematocrit or in studies where blood rheology is central to the clinical question.

Haematocrit influences blood viscosity and may affect coronary computational fluid dynamics (CFD). However, previous studies examined broad or pathological haematocrit ranges, and it remains unclear whether female-specific haematocrit variations within the physiological range produce meaningful changes in coronary haemodynamics. 15 female coronaries were analysed, including healthy arteries and diseased models with mild, moderate and severe stenosis. A haematocrit-dependent Carreau-Yasuda model was developed. Coronary CFD simulations were performed using the standard rheology model and a female-specific haematocrit-based model (40%). Time-averaged endothelial shear stress (TAESS), ESS gradient (ESSG), temporal shear variation index (TSVI), helicity intensity, and low/high TAESS exposure were compared across coronary trees, arterial segments, bifurcations, stenosed vessels and corresponding narrowed regions. The female-specific model produced statistically significant differences from the standard model across all haemodynamic metrics and coronary regions ($p < 0.05$). However, the absolute differences were small, indicating a limited haemodynamic impact. Bland-Altman analysis showed narrow biases and limits of agreement. Linear regression demonstrated significant associations between inter-model differences and haemodynamic magnitude for TAESS, ESSG, helicity intensity, and adverse TAESS exposure, but the slopes were small. Similar findings were observed in stenosed arteries, where both models captured comparable flow disturbances across stenosis severities. Female-specific haematocrit variation within the physiological range is computationally detectable but haemodynamically negligible in coronary CFD. A standard rheology model is therefore likely sufficient for most coronary CFD studies, while personalised haematocrit modelling is more relevant for patients with abnormal haematocrit or rheology-focused studies.